\documentstyle[prl,aps,epsfig,multicol,amssymb]{revtex}
\begin{document}
\draft

\title{Disorder induced cross-over effects at quantum critical points}
\author{Enrico Carlon$^{1}$ P\'eter Lajk\'o$^{2}$, and 
Ferenc Igl\'oi$^{3,2,4}$}
\address{
$^1$ INFM, Dipartimento di Fisica, Universit\'a di Padova, 
I-35131 Padova, Italy\\
$^2$ Institute for Theoretical Physics,
Szeged University, H-6720 Szeged, Hungary\\
$^3$ Research Institute for Solid State Physics and Optics, 
H-1525 Budapest, P.O.Box 49, Hungary\\
$^4$ Centre de Recherches sur les Tr\`es Basses Temp\'eratures, B. P. 166,
F-38042 Grenoble, France
}

\date{\today}
\maketitle

\begin{abstract}
Critical properties of quantum spin chains with varying degrees of disorder
are studied at zero temperature by analytical and extensive density matrix 
renormalization methods.
Generally the phase diagram is found to contain three phases. The weak 
disorder regime, where the critical behavior is controlled by the fixed 
points of the pure system and the strong disorder regime, which is attracted 
by an infinite randomness fixed point, are separated by an intermediate
disorder regime, where dynamical scaling is anisotropic and the static and 
dynamical exponents are disorder dependent.
\end{abstract}

\pacs{05.50.+q, 64.60.Ak, 68.35.Rh} 

\newcommand{\bc}{\begin{center}}
\newcommand{\ec}{\end{center}}
\newcommand{\be}{\begin{equation}}
\newcommand{\ee}{\end{equation}}
\newcommand{\ba}{\begin{array}}
\newcommand{\ea}{\end{array}}
\newcommand{\beqn}{\begin{eqnarray}}
\newcommand{\eeqn}{\end{eqnarray}}
\begin{multicols}{2}
\narrowtext

Quantum phase transitions take
place at zero temperature due to quantum fluctuations by varying a quantum 
control parameter \cite{sachdev},
which could be either the density of impurities 
(for the metal-insulator transition\cite{met_ins} in three dimensions (3d)
or for the superconducting-insulator transition\cite{sup_ins} in 2d) 
or an external thermodynamical parameter
such as the strength of the magnetic field in quantum spin glasses\cite{qsgexp}
or the pressure in $BaVS_3$\cite{mihaly}.
The effect of quenched, i.e. time-independent disorder at quantum phase transitions
is the subject of recent intensive research\cite{bhatt}; 
we mention, for instance, the recent
experimental studies about doping
effects in quantum spin chains\cite{sp_chain} and ladders\cite{sp_ladder} and 
a theoretical explanation about the non-Fermi liquid behavior in U and Ce 
intermetallics\cite{nonF}.

From a theoretical point of view several analytical and numerical studies have been
performed mainly in low-dimensional systems\cite{bhatt}. In particular in one 
dimension many, presumably exact, results have been obtained by the application 
of a strong disorder renormalization group (SDRG) method, originally 
introduced by Ma, Dasgupta and Hu\cite{mdh}. For the random transverse-field 
Ising model Fisher\cite{fisher} has shown that at the critical point 
the probability distributions of the couplings and the transverse fields 
broaden without limits under renormalization, and as a consequence the SDRG 
becomes asymptotically exact at the so-called infinite randomness fixed point 
(IRFP).

In the IRFP dynamical scaling is strongly anisotropic, such that the 
relaxation time, $t_r$, and correlation length, $\xi$, are related as:
$\ln t_r \sim \sqrt{\xi}$.
Scaling of the static singular quantities can be conveniently described
by the magnetization scaling dimensions, $x$, (and similarly for surface 
spins by $x^s$) and by the critical exponent of the average correlation 
length, $\nu$, which are given by\cite{fisher}:
\be
x_{\rm IRFP}=\frac{3-\sqrt{5}}{4},\quad 
x_{\rm IRFP}^s=\frac{1}{2},\quad 
\nu_{\rm IRFP}=2\;.
\label{exp_RTIM}
\ee
The asymptotically exact SDRG analysis of the
random transverse-field Ising model
has been extended into the
off-critical region \cite{ijl01}, i.e. into the Griffiths phase\cite{griffiths}
and the dynamical exponent, $z$, connecting time- and length-scales as
$t_r \sim \xi^{z}$, has been analytically calculated.

The SDRG method has been applied for a series of random quantum systems at the
transition point \cite{mdh,fisherxx,senthil,hyman,monthus}.
Despite the fact, that in the absence of
disorder, these systems have different type of (spatial and temporal)
correlations, thus they are in different universality classes,
the SDRG analysis indicates that if the disorder is strong enough
all models are attracted by the IRFP of the 
random transverse-field Ising model.
In that respect quantum fluctuations seem to be irrelevant, so that solely 
the strong disorder effects dominate the critical behavior.

However the SDRG approach does not provide any information on the behavior of 
the system at weak disorder. In this limit one generally invokes the 
Harris criterion \cite{harris,chayes} according to which weak disorder is 
irrelevant if:
\be
\nu_0 > 2/d\;,
\label{harris_cr}
\ee
where $\nu_0$ is the correlation length critical exponent of the pure system.
If $\nu_0 < 2/d$, thus disorder is relevant the IRFP could be strongly 
attractive as for the random transverse-field Ising model,
while if $\nu_0 > 2/d$ only a finite amount of disorder may bring the system to the IRFP.

The aim of this Letter is to investigate in detail the weak to strong disorder
crossover behavior at a quantum critical point and to present a possible
common scenario. In view of the generality of this phenomenon and its possible
relevance in realistic quantum systems it is of importance to perform a
detailed study of the singular behavior of random quantum systems in the whole
range of strength of disorder.

As a subject of our study we have chosen two one-dimensional quantum models, 
the $q$-state quantum clock model (CM) and the quantum Ashkin-Teller (AT) 
model. Our selection is motivated by the following facts: i) both models are 
self-dual, thus the location of the critical point is exactly known. ii) The 
SDRG analysis can be performed for both models leading to IRFP behavior for 
strong disorder. iii) Critical properties of the pure models are known in 
details, in particular one can locate the part of the phase diagram, where the 
Harris irrelevance criterion is valid in the form of Eq.(\ref{harris_cr}).

We start to introduce the quantum CM\cite{QCM}, which is defined by the 
Hamiltonian:
\be
H_{CM}=-\sum_l \left[J_l \cos \frac{2\pi (s_l-s_{l+1})}{q}
+ \frac{h_l}{2} \left({\bf M}_l^+ + {\bf M}_l^-\right)\right]\;,
\label{HCM}
\ee
in terms of a discrete spin variable, $s_l=1,2,\dots,q$, at a lattice site, 
$l$, and ${\bf M}_l^{\pm} |s_l\rangle=|s_l \pm 1,{\rm mod}~q \rangle$ is a 
spin raising (lowering) operator.

Similarly the quantum AT model is defined by the Hamiltonian\cite{kohmoto}:
\beqn
H_{\rm AT} &=& -\sum_l J_l(\sigma_l^z \sigma_{l+1}^z + 
\tau_l^z \tau_{l+1}^z) -\sum_l h_l (\sigma_l^x+\tau_l^x) - 
\nonumber \\
&&\epsilon \sum_l (J_l \sigma_l^z \sigma_{l+1}^z \tau_l^z \tau_{l+1}^z 
+h_l \sigma_l^x \tau_l^x)\;,
\label{HAT}
\eeqn
in terms of two sets of Pauli matrices, $\sigma_l^{x,z}, \tau_l^{x,z}$. For
both models the couplings, $J_l$, and the transverse fields, $h_l$, are 
independent random variables, while for the AT model the coupling between 
the two Ising models, $\epsilon$, is disorder independent. 
Both models are self-dual, which amounts to the invariance of the Hamiltonians in  
Eqs.(\ref{HCM})  and (\ref{HAT}) under the transformation $J_l \leftrightarrow 
h_l$.

The pure CM, with $J_l=J$ and $h_l=h$, for $q>4$ has an extended critical 
phase at both sides of the self-duality point, which is located at $J=h$. 
According to our numerical results, obtained by the density matrix 
renormalization group (DMRG) method \cite{DMRGbook} the critical exponents 
of the $q=5$ model at the self-duality point are given by:
\be
x_{\rm CM}=0.105(5),\quad 
x_{\rm CM}^s=0.23(1).
\label{expCM}
\ee
Since the self-duality point is located in the middle of a critical phase $\nu_{CM}$ is
formally infinity, thus small disorder is irrelevant, according to Eq.(\ref{harris_cr}).

The pure quantum AT model is critical along the self-duality line for $-1<\epsilon \le 1$ 
with the critical exponents:
\be
x_{\rm AT}=\frac{1}{8},\quad 
x_{\rm AT}^s=\frac{\rm{arccos}(-\epsilon)}{\pi}, \quad 
\nu_{\rm AT}=\frac{2 x_{\rm AT}^s}{4 x_{\rm AT}^s -1}\;.
\label{expAT}
\ee
which are conjecturedly exact \cite{kohmoto,rittenberg}.
For $-1/\sqrt{2}<\epsilon< -1/2$ the correlation length exponent of the pure 
model, $\nu_{\rm AT}$, exceeds the value of 2, thus the Harris irrelevance 
criterion in Eq.(\ref{harris_cr}) sets in.
(We note that $\nu_{\rm AT}$ stays formally infinite in the whole region of
$-1<\epsilon<-1/\sqrt{2}$, which is the central part of the so called
{\it critical fan}\cite{kohmoto}.)

The SDRG transformation can be performed for both models leading to similar
decimation rules:
\be
\tilde{h_i}=\frac{h_i h_{i+1}}{J_i \kappa},\quad 
\tilde{J_i}=\frac{J_i J_{i+1}}{h_i \kappa} \;,
\label{deciCM}
\ee
where the first (second) equation refers to the elimination of a strong 
bond (field) of strength, $J_i$ ($h_i$). For the CM
$\kappa=[1-\cos(2\pi/q)]/[1+\delta_{2,q}]$ \cite{senthil} and $q$ does 
not renormalize under the transformation. On the other hand for the AT 
model, where, $\kappa=1+\epsilon$, the coupling, $\epsilon$, enters into 
the renormalization as $\tilde{\epsilon}=\epsilon^2(1+\epsilon)/2$.
The RG equations in Eq.(\ref{deciCM}) are very similar to that of the 
random transverse-field Ising chain, which is recovered with $\kappa=1$. 
Both systems will scale to IRFP of that model for strong enough initial 
disorder and for $\kappa>0$.
Note, however, that for $0<\kappa<1$, i.e. for $q>4$ for the CM and 
$-1<\epsilon<0$ for the AT model, in some cases the generated new 
couplings/fields could be larger than the decimated ones, therefore the 
SDRG is not valid for weak disorder.

To see in detail the weak-to-strong disorder cross-over phenomena 
we perform a parallel numerical study about the critical 
behavior of the two models with varying degree of disorder. We use 
the DMRG method and calculate the average magnetization profiles, $m_l$,
fixing the spins at one boundary, $s_1=1$, and $\sigma_1^z=\tau_1^z=1$, 
respectively, whereas the spins at the other end of the chains, at $l=L$, 
are free. In particular we consider the scaling of the magnetization for spins 
at the bulk, when $l=L/2$ and $m_b \equiv m_{L/2}$, and that at the surface, 
when $m_s \equiv m_{L}$:
\be
m_b \sim L^{-x},\quad m_s \sim L^{-x^s}\;,
\label{magn_scale}
\ee
from which the scaling dimensions, $x$ and $x^s$ are calculated.

Useful information about the dynamical exponent, $z$, can be obtained from the
distribution function of the surface magnetization, which has the limiting
behavior \cite{ijl01}:
\be
P(\ln m_s) \sim m_s^{1/z'},\quad m_s \to 0\;.
\label{distr_ms}
\ee
The scaling exponent, $z'$, in Eq.(\ref{distr_ms}) should be compared with that
of due to pure quantum fluctuations, $z_q=1$, and then $z={\rm max}(1,z')$.

For both models we perform the calculations at the self-duality point, which corresponds 
to the critical point of the pure systems, as well as, according to the SDRG 
analysis, to the fixed-points of the strongly disordered systems. By choosing 
the same distribution function, $P(y){\rm d}y$, of the couplings 
$J_l$ and that of the fields $h_l$, we ensure to stay 
at the self-duality point and use the parametrization:
\be
P(y)=\Delta y^{-1+1/\Delta},\quad 0\le y \le 1\,\quad 
0< \Delta < \infty\;.
\label{distr_y}
\ee
One can see here that $\Delta$ plays the role of the strength of disorder: 
the pure system is recovered as $\Delta \to 0$, whereas as $\Delta \to \infty$ 
the distribution corresponds to that in the IRFP \cite{fisher}.

In the actual calculations we used $q=5$ for the CM while for the AT model we 
took $\epsilon=-0.75$. The bulk and surface magnetizations 
are calculated on relatively large finite systems up to $L=32$ and 10000 
disorder realizations were used, except for the largest sizes of the AT model, 
when we have some 1000 samples. For stronger disorder, i.e. for larger 
$\Delta$ the energy gap in a typical sample is decreasing, therefore to 
maintain sufficient numerical accuracy of the code one needs to keep relatively more
states in the DMRG procedure. As a consequence the computational demand of the
calculation is strongly increasing with $\Delta$, therefore we have restricted
ourselves to calculate only at few points in the strongly disordered regime.
Generically the DMRG algorithm performs rather well for disordered systems, as 
it is known also from other several examples \cite{DMRGdis,ijl01}.

\begin{figure}[h]
\centerline{\psfig{file=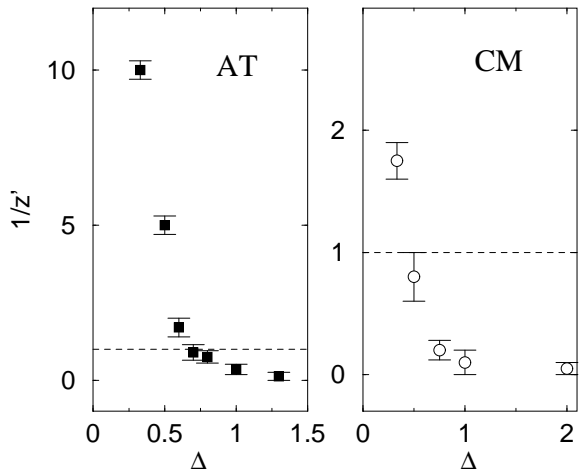,height=6.4cm}}
\vskip 0.08truecm
\caption{
Plots of $1/z'$ vs. $\Delta$ for the random AT model with $\epsilon=-0.75$
($\scriptstyle\blacksquare$) and for the CM with $q=5$ ($\circ$). 
The limiting values of $\Delta_0$ and 
$\Delta_{\infty}$, where $1/z'=1$ and $1/z'=0$, respectively are model 
dependent. The error bars in the figure are due to finite size effects 
and disorder averaging.}
\label{FIG01}
\end{figure}

Figure \ref{FIG01} shows the inverse of the disorder induced dynamical exponent, $1/z'$,
as calculated from Eq.(\ref{distr_ms}) for different values 
of the disorder strength, $\Delta$. In each cases $1/z'$ is found to be
a monotonically decreasing function of $\Delta$, such that $1/z'$ is lowered 
below the critical limit of $1/z'=1/z_q=1$ for a model dependent disorder 
strength, $\Delta_0$. Then, by further increase of $\Delta$ we obtain for 
the true dynamical exponent of the model, $z=z'$, to diverge, thus $1/z=0$ 
in the range of $\Delta_{\infty}<\Delta<\infty$, where the upper disorder 
limit, $\Delta_{\infty}$ is also model dependent.

\begin{figure}[h]
\centerline{\psfig{file=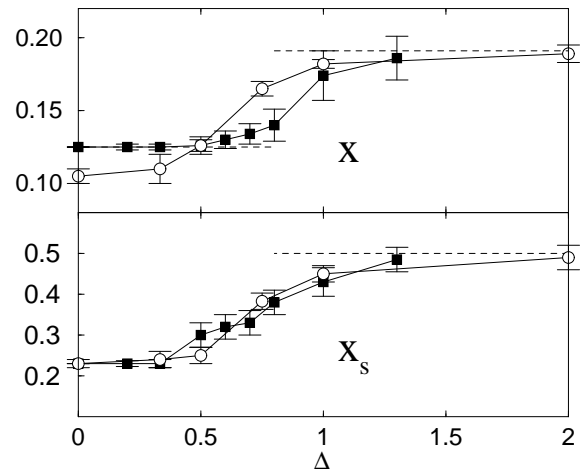,height=6.4cm}}
\vskip 0.08truecm
\caption{Scaling exponents of the average bulk and surface magnetization as 
a function of the strength of disorder for the $q=5$ CM
($\circ$) and for the AT model with $\epsilon=-0.75$
($\scriptstyle\blacksquare$). The limiting values of 
the exponents, corresponding to that of the IRFP are 
denoted by dashed lines.}
\label{FIG02} 
\end{figure}

Next in Fig. \ref{FIG02} the scaling exponents of the average magnetization, 
$x$ and $x^s$, as calculated from Eq.(\ref{magn_scale}) are shown for different 
strengths of disorder. Apparently, for weak enough disorder both the bulk and 
the surface exponents stay constant keeping the corresponding pure system 
value, which is in accordance with the Harris irrelevance criterion. Then, by 
increasing the strength of disorder over a limiting value, which approximately 
corresponds to $\Delta_0$, found in Fig.1, the exponents start to increase.
This variation is continuing until the exponents reach their IRFP values, as 
given in Eq.(\ref{exp_RTIM}), then they stay constant for further increase of 
the disorder. This second limiting value of the strength of the disorder is
approximately consistent with $\Delta_{\infty}$, what can be identified in 
Fig. \ref{FIG01} for the different models. Note, that the average behavior of
the systems, as presented in Fig. \ref{FIG02}, is dominated by rare 
realizations, which are samples with $m_b(m_s)=O(1)$, but which occurs with 
a very low probability. Therefore to obtain accurate averages one have to 
calculate on a large amount of samples, and the accuracy of the calculation 
turned out to be larger for bulk exponents than that of the surface ones.

From the numerical and the SDRG results obtained for the two models we can
extract a common route from the weak to strong disorder crossover, which is
expected to be generally valid
for a class of random critical quantum systems. This scenario is summarized 
in a generic RG phase diagram depicted in Fig. 3 in terms of the strengths of 
disorder, $\Delta$, and a model dependent parameter, say $\omega$, which is used
to characterize points in a critical line, 
(c.f. $\omega=\epsilon$ for the AT model, 
$\omega=4-q$ for the CM and $\omega=q/2-1$ for the quantum Potts model). For 
$\omega \ge 0$ the system is in the strong disorder (SD) phase, where
the IRFP is strongly attractive. In the other part of the phase diagram, 
for $\omega < 0$, there are two more phases: the weak disorder (WD) and the 
intermediate disorder (ID) regions. Here the SD part of the phase diagram is 
attracted by the IRFP, while the critical behavior in the WD regime is 
governed by the fixed points of the pure system, which are located at 
$\Delta=0$. In the ID regime, where there is a competition between quantum 
fluctuations and disorder effects, dynamical scaling is anisotropic, 
$1<z<\infty$, and the static and dynamical critical exponents are disorder 
dependent. At the boundaries of the ID region there are $1/z=0$ and $1/z=1$, 
respectively, while at $\Delta=0$, i.e. in the pure system limit they are
at $\omega=0$ and $\omega=\omega_0$. In the latter case for the pure model
with a parameter $\omega_0$ the Harris criterion in Eq.(\ref{harris_cr}) is 
saturated, i.e. $\nu_0(\omega_0)=2/d$.
\begin{figure}[h]
\centerline{\psfig{file=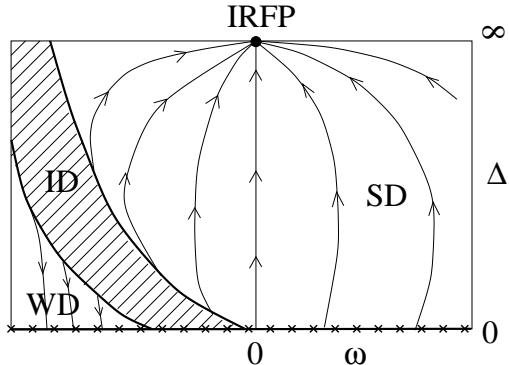,height=4.8cm}}
\vskip 0.08truecm
\caption{Schematic RG phase diagram of a quantum spin system having a 
critical line parametrized by $\omega$, in the presence of disorder of 
strength, $\Delta$. For details see the text.}
\label{FIG03}
\end{figure}
We believe that the above scenario is relevant for some higher dimensional systems and 
therefore could have experimental relevance, too. As an example we mention 
the two dimensional random transverse-field Ising model
where by a numerical implementation of the SDRG method an IRFP 
is found\cite{2drg}, which is attractive for strong disorder both for 
spin-glasses (SG-s) and random ferromagnets (FM-s). In numerical studies,
however, a finite dynamical exponent, $z$, is obtained for the SG model
\cite{sg}, while for the random FM $z$ is divergent at the critical point
\cite{2dmc}. A possible explanation of this controversy is that the studied 
SG is still in the ID region, whereas the random FM is in the SD phase. 
Further examples could be provided by random antiferromagnetic spin systems:
chains, ladders and 2d models, where often conventional fixed-point behavior 
is obtained \cite{hyman,monthus,mlri}.

The authors are grateful to R. Melin, H. Rieger and L. Turban for 
useful discussions. This work has been supported by a German-Hungarian 
exchange program (DAAD-M\"OB), by the Hungarian National Research Fund under 
grant No OTKA TO23642, TO25139, TO34183, F26004 and MO28418 and by the 
Ministry of Education under grant No FKFP 87/2001.

\end{multicols}
\widetext

\begin{references}
\bibitem{sachdev}
	For a review see: S. Sachdev: {\it Quantum Phase Transitions}, 
        Cambridge University Press, Cambridge (1999).

\bibitem{met_ins}
	P.~A.~Lee and T.~V.~Ramakrishnan, Rev.  Mod. Phys. {\bf 57}, 287 (1985).

\bibitem{sup_ins}
	Y. lin {\it et al.} Phys. Rev. Lett. {\bf 67}, 2068 (1991).

\bibitem{qsgexp} 
        W. Wu {\it et al.}
	Phys. Rev. Lett. {\bf 67}, 2076 (1991);
    	W. Wu, D. Bitko, T. F. Rosenbaum and G. Aeppli,
	Phys. Rev. Lett. {\bf 71}, 1919 (1993).

\bibitem{mihaly}
        L. Forr\'o {\it et al.}
        Phys. Rev. Lett. {\bf 85}, 1938 (2000).

\bibitem{bhatt}
	For a review, see R. N. Bhatt, in {\it Spin Glasses and Random Fields},
	edited by A. P. Young (World Scientific, Singapore, 1998), p. 225. 

\bibitem{sp_chain}
	G. Theodorou and M. H. Cohen, Phys. Rev. Lett. {\bf 37}, 1014 (1976);
	Y. Uchiyama {\it et al.}, Phys. Rev. Lett. {\bf 83}, 632 (1999); 
	E. Janod {\it et al.}, cond-mat/0103379.

\bibitem{sp_ladder}
	M. Azuma, M. Takano, R. S. Eccleston, cond-mat/9706170

\bibitem{nonF}
	A. H. Castro Neto and B. A. Jones, cond-mat/0003085.

\bibitem{mdh}
	S. K. Ma, C. Dasgupta and C.-K. Hu, Phys. Rev. Lett. {\bf 43}, 1434 
        (1979); C. Dasgupta and S. K. Ma, Phys. Rev. B{\bf 22}, 1305 (1980).

\bibitem{fisher}
        D. S. Fisher, Phys. Rev. Lett. {\bf 69}, 534 (1992); 
        Phys. Rev. B {\bf 51}, 6411 (1995).

\bibitem{ijl01}
	F. Igl\'oi, R. Juh\'asz and P. Lajk\'o, Phys. Rev. Lett. {\bf 86}, 
        1343 (2001).

\bibitem{griffiths}
        R. B. Griffiths, Phys. Rev. Lett. {\bf 23}, 17 (1969).

\bibitem{fisherxx}
        D. S. Fisher, Phys. Rev. B {\bf 50}, 3799 (1994).

\bibitem{senthil}
	T. Senthil and S. N. Majumdar, Phys. Rev. Lett. {\bf 76}, 3001 (1996).

\bibitem{hyman}
	R. A. Hyman and K. Yang, Phys. Rev. Lett. {\bf 78}, 1783 (1997).

\bibitem{monthus}
	C. Monthus, O. Golinelli and Th. Jolicoeur, Phys. Rev. Lett. {\bf 79}, 
        3254 (1997).

\bibitem{harris}
	A. B. Harris, J. Phys. C {\bf 7}, 1671 (1974).

\bibitem{chayes}
	J. T. Chayes, L. Chayes, D. S. Fisher and T. Spencer,
        Phys. Rev. Lett. {\bf 57}, 299 (1986).

\bibitem{QCM}
	J. Jose, L. Kadanoff, S. Kirkpatrick, and D. R. Nelson, Phys. Rev. B 16, 1217 (1977);
	S. Elitzur, R. Pearson, and J. Shigemitsu, Phys. Rev. D 19, 3698 (1979). 
\bibitem{kohmoto}
	J. Ashkin and E. Teller, Phys. Rev. {\bf 64}, 178 (1943);
	M. Kohmoto, M. den Nijs and L.P. Kadanoff, Phys. Rev. B {\bf 24}, 
        5229 (1981).
 
\bibitem{DMRGbook} See e.q. in {\it Density Matrix Renormalization: A New Numerical 
Method in Physics}, edited by I. Peschel, X. Wang, M. Kaulke, and K. Hallberg
(Springer, Berlin, 1999).

\bibitem{rittenberg}
	G. von Gehlen and V. Rittenberg, J. Phys. {\bf A} 19, L1039 (1986); {\bf 20}, 227 (1987). 

\bibitem{DMRGdis} E. Carlon, C. Chatelain and B. Berche, \prb {\bf 60}, 
                  12974 (1999); S. Rapsch, U. Sch\"ollwock and W. Zwerger, 
                  Europhys. Lett. {\bf 46}, 559 (1999).

\bibitem{2drg}
        O. Motrunich, S.-C. Mau, D.A. Huse and D.S. Fisher, Phys. Rev. B
	{\bf 61}, 1160 (2000);
        Y-C. Lin, N. Kawashima, F. Igl\'oi and H. Rieger, Prog. Th. Phys. 
	(Suppl.) {\bf 138}, 470 (2000).

\bibitem{sg}
	M. Guo, R.N. Bhatt and D.A. Huse, Phys. Rev. Lett. {\bf 72}, 4137 
	(1994);
	H. Rieger and A.P. Young, Phys. Rev. Lett. {\bf 72}, 4141 (1994).

\bibitem{2dmc}
        C. Pich, A.P. Young, H. Rieger and N. Kawashima, Phys. Rev. Lett. 
	{\bf 81}, 5916 (1998).

\bibitem{mlri}
	R. Melin, Y-C. Lin, H. Rieger and F. Igl\'oi (unpublished).

\end{references}
\end{document}